\documentclass[showpacs,preprint,aps]{revtex4}
\usepackage{graphicx}
\usepackage{dcolumn}
\usepackage{bm}
\usepackage{amssymb} 
\usepackage{amsmath}
\begin{document}
\setcounter{page}{1}
\title 
{Orbiting phenomena in black hole scattering}
\author
{D. Batic$^1$, N. G. Kelkar$^2$ and M. Nowakowski$^2$}
\affiliation{ 
$^1$ Dept. of Mathematics, Univ. of West Indies, Kingston 6, Jamaica\\
$^2$ Dept. de Fisica, Universidad de los Andes,
Cra.1E No.18A-10, Bogota, Colombia}
\begin{abstract}
Rainbow, glory and orbiting scattering are usually described by the 
properties of the classical deflection function related to the real part 
of the quantum mechanical scattering phase shift or by the diffractive 
pattern of the quantum mechanical cross sections. Here we show that the 
case of orbiting scattering of massless spin 0, 1 and 2 particles 
from Schwarzschild black holes can be characterized by a 
sudden rise in $|R_l|^2$ at a 
critical angular momentum $l_C$, which we show corresponds to the 
unstable circular orbits of these particles. 
For the cases, $s =0, 2$, we attempt a 
new interpretation of the Regge-Wheeler potential by identifying 
the quantum
mechanical corrections to the effective potential of massless
particles. We probe into the black
hole scattering by using numerical and semi-analytical methods which give
very good agreements with the exact numerical 
results. The limitations of previously 
used approximations as compared to the exact and semi-analytical results 
are discussed.  
\end{abstract}
\maketitle 

\section{Introduction}
Over the last four decades, the physics of particle scattering from different 
kinds of black holes was one of the most active topics of strong 
gravitational fields. Apart from the quasinormal modes which, in principle,  
can be identified as the poles of the corresponding 
black hole scattering matrix, the behaviour of the cross section with respect 
to the scattering angle is one of the most interesting features in this area.
The key issues around which the physics of black hole scattering centers 
are related to phenomena such as glory, orbiting (or spiralling), rainbow 
and super-radiant scattering. One should note that these features 
are observed experimentally 
and studied extensively theoretically in nucleus-nucleus scattering. 
It is gratifying to see similar phenomena occurring in black hole 
scattering \cite{Matzners}. 
In the pioneering papers of Ford and Wheeler \cite{ford12}, using 
semi-classical arguments, a connection between the classical rainbow, glory 
and orbiting phenomena was made with their quantum mechanical counterparts. 
Starting with a procedure as proposed by Mott and Massey \cite{mottmass} 
which gave the classical cross section $\sigma_{cl} = \sigma_{semicl}$, the 
authors noticed that the classical deflection function can be written in 
terms of the quantum mechanical scattering phase shift as 
$\Theta (l(b)) \, =\, 2 d [\Re e \delta_l]/dl$, where $l$ is the angular 
momentum and $b$ the impact parameter. 
For particles scattered into the solid angle $\Omega(\theta,\phi)$, the
cross sectional area can be written as, $d\sigma = b \, db\, d\phi$, so
that the differential cross section is proportional to 
$|db/d\theta|/\sin(\theta)$. 
The classical angular momentum is $L = b p$. Going over to the quantum mechanical case,
if we now consider
the de Broglie wavelength $\lambda = k ^{-1}$ of the particle to be
small compared to the range of the force, we may write, $b = L/p \simeq 
\sqrt{l(l+1)}/k$. In fact, while working with semi-classical approximations
such as the WKB, one must replace $l(l+1) \to (l + 1/2)^2$.
In the article of Ford and Wheeler, the authors thus obtain the differential
cross section proportional to $(d\Theta/dl)/\sin(\theta)$, 
which is divergent when 
the scattering angle $\theta$ is $0$ or $\pi$. 
Glory is characterized by $\Theta(l)$ passing with finite slope 
through $0, \pm\pi$ etc and 
rainbow by the maxima or minima in $\Theta(l)$. If the deflection function 
displays a singularity at a certain critical value $l_C$ of $l$, 
they show that $\Theta(l)$ will vary logarithmically near $l = l_C$. 
For values of $l$ below and above $l_C$ the particle would have spiralling 
trajectories and $l = l_C$ would give the limit of an unstable circular 
orbit. For a particle incident with energy $E$ and for an effective 
potential, $V_{eff}$, which is the sum of an actual interaction potential 
and a centrifugal term, the condition corresponding to such an orbit was shown 
to be $V_{eff}(r_C, l_C) = E$ (which as we shall see later is 
$V_{eff}(r_C, l_C) = \omega^2$ for 
massless particles with $E^2 = \omega^2$), 
where $r_C$ is the position of the maximum 
of the effective potential and $E$ the available energy. 
Quantum mechanically, one does not get divergent cross sections. One rather 
observes peaks in the cross sections as a function of angle in the 
backward directions for glory and orbiting scattering. 

Based on the works of Ford and Wheeler, 
the characterization of the above phenomena in black hole scattering is 
mostly done in literature 
by noting the behaviour of the real part of the scattering 
phase shift or by looking at the oscillating patterns in the cross sections 
at backward angles. In the present work, we relate the phenomenon of 
orbiting scattering to the imaginary part of the scattering phase shift. 
To be specific we evaluate the reflection coefficient in black hole 
scattering from Schwarzschild black holes by solving the corresponding 
Riccati equation numerically. Finding a sudden rise of the reflection 
coefficient, $|R_l|^2 = exp(-4\delta_l^I)$ plotted as a function of $l$, 
at a certain critical value of $l$ for different energies and 
different spins ($s = 0, 1, 2$) of the massless scattering particles, we 
show that this critical $l$ is nothing but the $l_C$ corresponding to the 
unstable circular orbit. We also find that the normalized $|R_l|^2$ always 
passes through a value of $1/2$ at the critical value $l_C$. 

The present work differs from others \cite{Matzners} in view of the points 
mentioned above. The reflection coefficient is evaluated exactly 
using the variable amplitude method as compared to approximate 
calculations (third reference in \cite{Matzners}), \cite{fabbri}. 
We also find a potential proportional to $\cosh^{-2}(\alpha x -\beta)$
which gives remarkably good results when compared with the numerical ones.
For this potential the transmission coefficients can be found analytically.
Parametrizing the $\cosh^{-2}$ potential to fit the Regge-Wheeler potential, 
we find semi-analytical results for black hole scattering.
Even though the scattering off black holes is a widely explored area, 
not much attention has been paid to the orbiting phenomenon (mostly
the glory and rainbow effects have been discussed). Here we
supplement the existing literature by a detailed study of how the 
existence of a classical
orbit gets reflected in the quantum mechanical 
expressions of the scattering of a massless particle from a black hole.
This leads to a conjecture regarding quantum corrections to the
classical effective potential for massless particles.

In the next section, 
we provide briefly the formalism for black hole scattering in general 
and go on to discuss the critical parameters relevant to orbiting 
and glory scattering. 
In section III, we discuss the calculation of the reflection coefficient 
and present results regarding its connection to orbiting scattering. 
We also present a conjecture related to these phenomena. 
In section IV we discuss how the Regge-Wheeler potential can be
re-interpreted as an effective potential plus quantum corrections
proportional $\hbar$. 
In section V we summarize our findings. 

\section{Black hole scattering}
We start by presenting some generalities in black hole scattering. 
Consider the propagation of a massless scalar field
$\phi=\phi(t,r,\vartheta,\varphi)$ governed by the wave equation
$g^{\mu\nu}\nabla_\mu\nabla_\nu\phi=0$ where $g_{\mu\nu}$ denotes
a static, spherically symmetric black hole metric whose line
element is
\[
ds^2=f(r)dt^2-\frac{dr^2}{f(r)}-r^2(d\vartheta^2+\sin^2{\vartheta}d\varphi^2)
.
\]
Using the following ansatz
\begin{equation}\label{ansatz}
\phi(t,r,\vartheta,\varphi)=e^{i\omega t}\frac{1}{r}~\psi_{n \ell
\omega}(r)Y_{\ell m}(\vartheta,\varphi),\quad {\rm{Re}(\omega)>0}
\end{equation}
it is standard \cite{BHscatt} to reduce the above equation 
to a Schr\"{o}dinger-like equation for the radial part
\begin{equation}\label{radial}
\left[-\frac{d^2}{dr_{*}^2}+V(r)\right]\psi_{n \ell \omega
}=\omega^2\psi_{n \ell \omega },
\end{equation}
where in principle $\omega \equiv \omega_n$, but we shall drop 
the subscript for convenience in what follows. 
Moreover, $V(r)=f(r)U(r)$ (with the form of $f(r)$ depending on the 
metric under consideration) and
\begin{equation}\label{blackpotUr} 
U(r)=\frac{l(l+1)}{r^2}+\frac{f^{'}(r)}{r}.
\end{equation}
Here, a prime denotes differentiation with respect to $r$ whereas
$r_{*}$ is a tortoise coordinate defined through
\[
\frac{dr_{*}}{dr}=f(r)^{-1}.
\]

\subsection{Scattering from a Schwarzschild black hole}
In case of the Schwarzschild metric which we shall consider in the 
present work, $f(r)=1-2M/r$ where $M$ is the mass of the black hole
and the tortoise coordinate is given by
\[
r_{*}=r+2M \ln{\biggl(\frac{r}{2M}-1\biggr)},\quad r>2M.
\]
Eq. (\ref{radial}) can also be obtained for other spins. The corresponding 
Regge-Wheeler potential for spins $s = 0, 1, 2$ is given as \cite{visser}
\begin{equation}\label{blackpot}
V(r(r_*)) = \biggl (1 - {2M \over r}\biggr)\, \biggl[ {l(l+1)\over r^2} 
+ {2M (1 - s^2)\over r^3} \biggr ]
\end{equation}
where, $l \ge s$. 
Note that as we move from $r=r_0=2M$ at the event horizon to $r = \infty$, 
the tortoise coordinate varies from $-\infty$ to $\infty$. 
Since the scattering problem with the radial coordinate $r$ in three 
dimensions (3D) gets mapped into a one-dimensional (1D) 
one with the $r_*$ coordinate, 
the Schr\"odinger-like equation (\ref{radial}) in black hole scattering
can be solved using standard techniques for 
1D tunneling in quantum mechanics.
The asymptotic solutions of the Schr\"odinger
equation (\ref{radial}) are
$$\psi (r_*)\, =\, A(\omega) e^{+i\omega r_*} \, +\, B(\omega) 
e^{-i\omega r_*}, \,\,\,\, r_* \to - \infty,$$
$$\psi (r_*)\, =\, C(\omega) e^{+i\omega r_*} \, 
+\, D(\omega) e^{-i\omega r_*}, \,\,\,\, r_* \to + \infty .$$
For waves incident on the black hole from the right (i.e. $+\infty$) we have 
$B(\omega)=0$, the reflection amplitude
$R(\omega) = D(\omega)/C(\omega)$
and the transmission amplitude
$T(\omega) = A(\omega)/C(\omega)$, so that
$$\psi (r_*) \, =\,  T(\omega) e^{i\omega r_*}, \,\,\,\, r_* \to - \infty,$$
$$\psi (r_*)\, =\, e^{i\omega r_*} \, +\, R(\omega)  e^{-i\omega r_*}, 
\,\,\,\, r_* \to + \infty. $$

\subsection{Critical parameters for black hole orbiting} 
An anomalous large angle scattering, called ALAS, was observed often in 
nuclear reactions between $\alpha$-like nuclei such as 
$^{12}$C-$^{16}$O, $^{16}$O-$^{28}$Si etc \cite{braun} 
and has been attributed to the orbiting mechanism in scattering.
The origin of this mechanism can be traced back to classical dynamics, 
where a particle approaching the potential center can get trapped 
in a circular orbit of radius $r_0$ if its energy equals the maximum 
of the effective potential at $r_0$. Ford and Wheeler \cite{ford12} 
found the connection of this phenomenon with the classical deflection 
function which becomes singular at a critical value of the angular 
momentum for which the circular orbit can exist and leads to divergent 
cross sections. 
The analogous effect in quantum mechanical scattering corresponds to the 
appearance of a diffraction pattern (or peaks) 
in the scattering cross section in the backward direction. 

\subsubsection{Deflection function in glory and orbiting}
It was shown in \cite{ford12} that as long as the classical deflection 
function $\Theta(l)$ remains between $0$ and $\pm \pi$, the semi-classical 
cross section can be entirely described by the classical cross section. 
If the deflection function passes smoothly through $0$ or $\pm \pi$, it 
leads to the phenomenon named glory. Though classically it corresponds to 
a singularity in the cross section, quantum mechanically one expects only a 
prominent peak in the cross section. 
Ford and Wheeler related the deflection function to the 
real part of the quantum mechanical scattering phase shift. 
Detailed discussions on the topic 
can be found in \cite{ford12} and \cite{nussen}. Here we directly state their 
conclusion, namely, 
\begin{equation}\label{deflphase}
\Theta (l) \, =\, 2 {d \delta_l^R \over dl}
\end{equation}
connecting the deflection function with the real part of the phase shift. 
There exists a critical value $l_g$ corresponding to backward 
glory scattering. The deflection function at backward 
angles can be approximated as 
\begin{equation}
\Theta(l) = \pi \,+ \,a\, (l - l_g)\, .
\end{equation}
Orbiting occurs when the effective potential as a function of the radial 
coordinate $r$ possesses for some angular momentum $l_C$, a relative maximum 
equal to the available energy. 

For massive particles 
in classical General Relativity (GR) this means $V_{eff}^{m \neq 0} (r_C, \ell_C) = E$. 
$V_{eff}$ enters the geodesic equation in the form $\dot{r}^2/2 + V_{eff} = const$.  
For massless particles, the same condition with $V_{eff}$ from the geodesic equation of motion is 
\begin{equation}\label{veffgeo} 
\biggl ( {d V_{eff}(\ell_C) \over dr}\biggr )_{r_C}\, =\, 0\,,\, \, \, \, \, 
V_{eff}(r_C,\ell_C) = \omega^2\, , 
\end{equation}
with 
\begin{equation}
V_{eff}(r) = {\ell^2 \over 2 r^2} - {M \ell^2 \over  r^3} \, . 
\end{equation}
where $\ell$ has the dimension of angular momentum per mass which makes 
$V_{eff}$ dimensionless. Replacing $\ell^2$ by $l (l + 1)$ we return back to 
the quantum mechanical picture. 
Note that part of $V$ (i.e. the first term of V(r) in (\ref{blackpot})) 
is proportional to $V_{eff}$ and in the case of $s = 1$, the entire 
$V (r)$ is proportional to  $V_{eff}$. We shall come back to this point later.
Under such a condition, the deflection function 
was shown to vary logarithmically 
\begin{eqnarray}
\Theta(l) \, =\, \theta_1 \, +\, b \ln \biggl( {l - l_C \over l_C}\biggr),
\quad l > l_C, 
\\ 
\nonumber
\Theta(l) \, =\, \theta_2 \, +\, 2b \ln \biggl( {l_C - l \over l_C}\biggr), 
\quad 0 \le l < l_C\, , 
\end{eqnarray}
where $\theta_1$, $\theta_2$ and $b$ are constants. 
The particle is expected to spiral below or above the barrier depending on 
the value of $l$ being greater 
or less than $l_C$, respectively. If $l = l_C$, the 
particle is trapped in a circular orbit and $\Theta(l)$ is singular. With 
$\Theta(l)$ being related to the real phase shift as in 
(\ref{deflphase}), one expects a steep 
jump down in the real part of the phase shift at the critical value of $l$. 

\subsubsection{Radius of the unstable orbits and critical $l$} 
In black hole scattering with $s=1$, Eq. (\ref{blackpot}) is proportional to the 
classical $V_{eff}$ from General Relativity. Here one expects 
an unstable photon orbit at $r_C = 3M$. 
Considering the potential $V(3M) = \omega^2$ with the semi-classical 
prescription of $l(l+1) \rightarrow (l + 1/2)^2$, 
it is easy to see that the critical value of the 
angular momentum $l$ is given by $l_C = (3 \sqrt{3}/2)\omega r_0 - 1/2$, 
where $r_0 = 2M$. 
If one uses ${l_C(l_C+1)}$ instead, one of course ends up with a 
quadratic equation for $l_C$. The two values of $l_C$ 
should however be quite close 
for large values of $l$. One could try to find the critical $l_C$ for the 
occurrence of circular orbits 
in the scattering of spin 0 and 2 particles in the same 
way too. Considering $V(r)$ in (\ref{blackpot}) at $r = 3M$ leads to 
\begin{equation}
l_C^{WKB} \, = \, \sqrt{{27\over 4} \omega^2 r_0^2 - {2\over 3}
(1 - s^2)} \, - \, {1\over 2}\, , 
\end{equation}
in the semi-classical approximation and 
\begin{equation}
l_C^{QM} \, =\, 
\sqrt{{27\over 4} \omega^2 r_0^2 - {2\over 3}
(1 - s^2)+ \frac{1}{4}}\,-\, {1 \over 2} \,
\end{equation}
quantum mechanically. 
In Table I we list the two sets of $l_C$ for different values of $\omega r_0$. 
As expected, the difference between the semi-classical $l_C^{WKB}$ and 
$l_C^{QM}$ is little for large values of $l$. Note that for $s=0$, there 
exists a critical $\omega r_0$ below which one cannot find a real $l_C$. 
\begin{table}[h]
\caption{\label{tab1}Critical values of $l$ obtained using $V(r = 3M)$.
The numbers outside parentheses correspond to $l_C^{WKB}$ and those inside 
to $l_C^{QM}$.}

\begin{tabular}{|l|l|l|l|}
\hline
$\omega \,r_0$ & $s = 0$ & $s = 1$ & $s = 2$ \\ \hline
0.5  &- &0.799 (0.892) &1.4203 (1.484) \\
1 &1.966 (2.017) &2.098 (2.145) & 2.458 (2.5) \\
2 &4.632 (4.656) & 4.696 (4.720) & 4.885 (4.908) \\
2.5  &5.944 (5.963) & 5.995 (6.014) & 6.147 (6.166) \\
3  & 7.251 (7.267) & 7.294 (7.310) & 7.422(7.437) \\ \hline
\end{tabular}
\end{table}

Instead of taking the value of the potential at $r = 3M$ which 
corresponds to the maximum in the classical effective potential, 
we now find $V(r_C)$ (where $r_C$ corresponds to the point where the 
maximum in $V(r)$ occurs) and use it to find the critical $l_C$.
Thus, setting $dV/dr = 0$, we find 
\begin{equation}
r_C = {3 r_0 \over 4} \biggl ( 1 - {(1 - s^2) \over L^2} \biggr )\, 
\biggl [ 1 \pm \sqrt{1 + {32 \over 9} {L^2 (1 - s^2) \over (L^2 - 1 + s^2)^2}}
\biggr ] 
\end{equation}
where, $L^2 = l_C (l_C + 1)$. Evaluating $V(r_C)$, 
one can now find $l_C$ by looking for the zeros of the function 
$r_0^2 V(r_C) - \omega^2 r_0^2$. In Table II, we list the values of critical 
$l_C$ evaluated as above for the scattering of spin 0, 1 and 2 particles. 
\begin{table}[h]
\caption{\label{tab2}Critical values of $l$ obtained using $V(r = r_C)$}
\begin{tabular}{|l|l|l|l|}
\hline
$\omega \,r_0$ & $s = 0$ & $s = 1$ & $s = 2$ \\ \hline
0.5  & 0.618& 0.892& 1.497\\
1 & 2.016 & 2.145 & 2.504 \\
2 & 4.656 & 4.720 & 4.909  \\
2.5  & 5.963 & 6.014  & 6.167  \\
3  & 7.267 & 7.310 & 7.438  \\ \hline
\end{tabular}
\end{table}
The reader will notice that apart from the value of $l_C$ which cannot be 
determined for $\omega r_0 = 0.5$ ($s =0$) in Table I, the remaining 
values are quite close to those in Table II. What appears at a first glance 
as a curious coincidence will be explained in the next section by analyzing 
the form of $V(r)$. 

In what follows, we shall present an exact numerical 
calculation of the reflection coefficient and study its behaviour as a 
function of $l$ in context with orbiting scattering. 
\section{Reflection coefficient and characterization of circular orbits}
In this section we will compare and discuss three different
methods to calculate the reflection coefficient: a semi-analytical result,
numerical results using the variable amplitude method and
the approximation of a rectangular barrier adjusted to the
problem of black hole scattering. We will see that
the semi-analytical result gives a very good
overall description of the problem.
The reflection amplitude in black hole scattering has also 
been calculated in literature \cite{fabbri} using semi-classical 
approximations. 
\subsection{Semi-analytical results}
Before we go over to the details of the
calculations of the reflection coefficient, 
let us briefly examine the nature of the potential in black hole scattering 
and what results one can expect. The potential in the Schwarzschild case as 
given in Eq.(\ref{blackpot}) is made up of two functions, namely, $U(r)$ 
(see Eq. (\ref{blackpotUr})) and $f(r)$, such that $V(r) = f(r) \, U(r)$. 
$U(r)$ consists of a $l$ dependent term which resembles the centrifugal 
barrier in standard problems in quantum mechanics. The form of $f(r)$ 
depends on the metric under consideration. In Fig. 1, we plot the potential 
as a function of the coordinate $r$ as well as $r_*$ and note the following 
features\\
(i) The potential looks very different when taken
as a function of $r$ or $r_{*}$.\\
(ii) If we plot the function $U(r)$ only, the steep rise of the centrifugal 
barrier is evident, however, only when plotted as a function of 
$r$ and not $r_*$. 
\\
(iii) Due to the presence of $f(r)$, the potential plotted as a function 
of $r_*$ resembles a Gaussian barrier and the centrifugal term is not 
explicitly seen in the shape of the potential. However, as evident from
Fig.2 the height of the potential rises with $l$.\\
\begin{figure}[h]
\includegraphics[width=7cm,height=10cm]{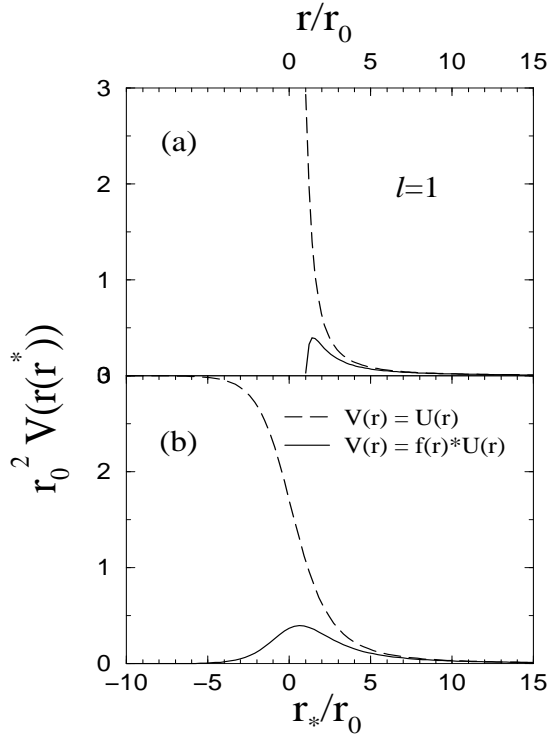}
\caption{\label{fig:eps1}
The Schwarzschild black hole potential for spin $s = 0$ and $l =1$. 
(a) Potential plotted as a function of the dimensionless 
coordinate $r/r_0$ and (b) $r_*/r_0$, where $r_*$ is the tortoise coordinate 
and $r_0$ the Schwarzschild radius.}
\end{figure}
\begin{figure}[h]
\includegraphics[width=9cm,height=8cm]{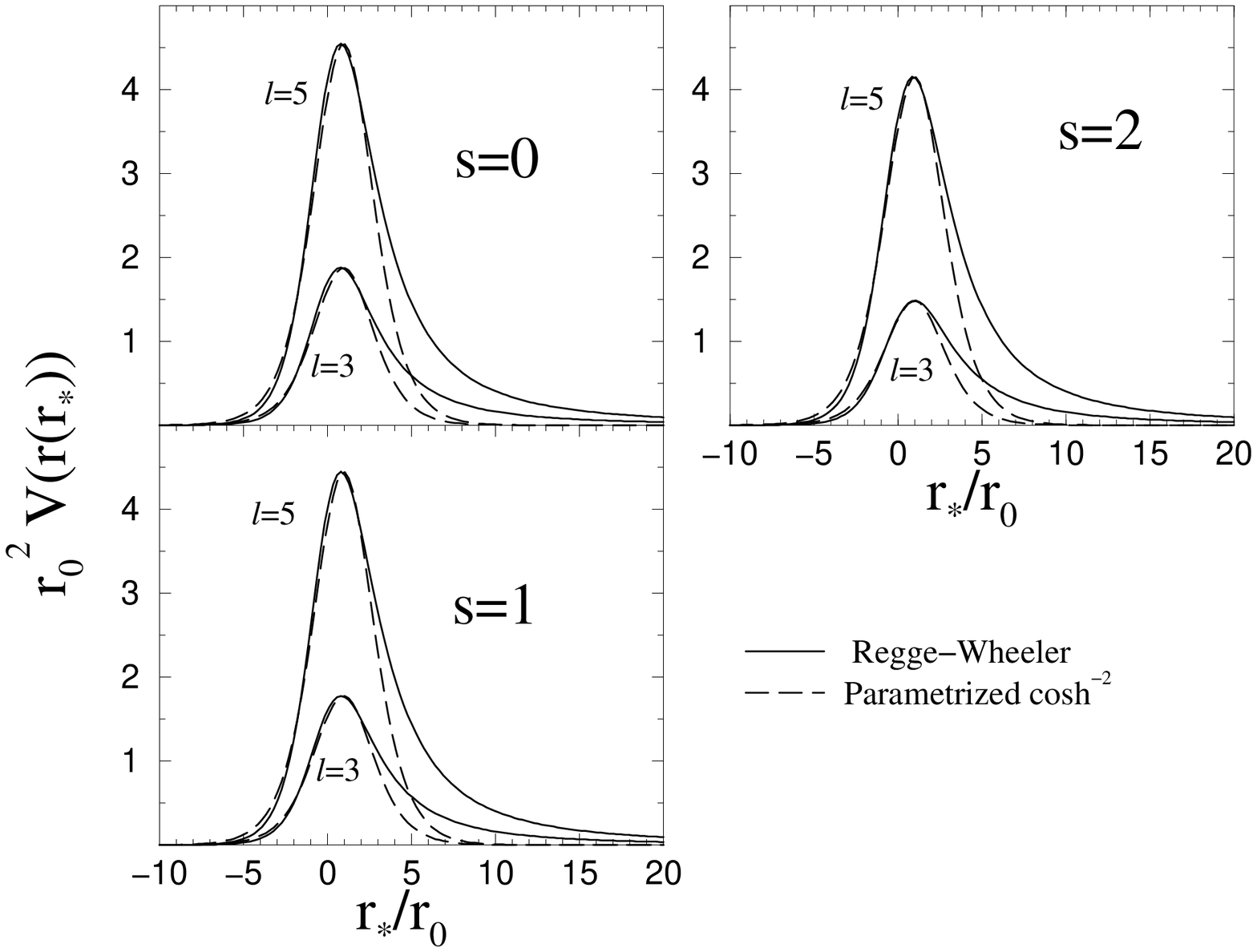}
\caption{\label{fig:eps2}
The Regge-Wheeler potential compared with the $\cosh^{-2}$ potential
from equation (\ref{cosh2}) for different spins and angular momentum. 
The discrepancy between the two cases is more prominent at smaller energies where the
Regge-Wheeler potential displays an asymmetric tail. Notice also
that increasing $l$ results in an increasing height.}
\end{figure}

New insights can be often gained by searching for analytical and
semi-analytical results. To this end we notice that the reflection
coefficients for the potential
\begin{equation} \label{cosh1}
U(x)=\frac{U_0}{\cosh^2(\alpha x)}
\end{equation}
(or modification of the above by a shift of the argument) can be
obtained analytically (we refer the reader for details to \cite{LL3}).
The relevance of this potential to our problem is its similarity
to the Regge-Wheeler potential in the tortoise coordinate. Indeed,
\begin{equation} \label{cosh2}
r_0^2V(r(r_*)) \approx \frac{r_0^2 V_0}{\cosh^2(r_*/ar_0 -b)}
\end{equation}
fits the Regge-Wheeler potential quite well for $s=0,1,2$ provided
we choose $a=2.4$, $b=0.4$ and the height $V_0$ to be the 
Regge-Wheeler potential at $r=3M$, i.e., $V_0=V(r=3M)$ which is different for
different choices of $s$ and $l$-dependent. 
For instance, in the cases $s=0,1,2$ we obtain
\begin{eqnarray}
r_0^2V_0=\frac{1}{27}\left [4l(l+1) + {8\over 3}\right]\, , \quad s = 0\\ \nonumber
r_0^2V_0=\frac{4}{27} 4l(l+1) \, ,\quad s = 1 \\ \nonumber 
r_0^2V_0=\frac{1}{27}\left [4l(l+1) -8\right] \, ,\quad s = 2
\end{eqnarray}
The comparison between
the Regge-Wheeler and the parametrized $\cosh^{-2}$ potential
is shown in Fig.2.  Evidently, one would expect
some quantitative agreement in both cases for the reflection
coefficient for tunneling at higher energies, i.e., where the two potentials
almost overlap. We will see that this is indeed the case.
To be able to use the analytical results from \cite{LL3}
we use: $k^2=\omega^2$, $\alpha=1/ar_0$ and $2mU_0=V_0$. This
gives the following transmission coefficients:
\begin{equation} \label{cosh3}
|T_l|^2=\frac{\sinh^2(a\pi\omega r_0)}{\sinh^2(a\pi \omega r_0)
+\cos^2(\pi/2\sqrt{(1-4V_0a^2r_0^2)})}
\end{equation}
if $4V_0a^2r_0^2 < 1$ and
\begin{equation} \label{cosh4}
|T_l|^2=\frac{\sinh^2(a\pi\omega r_0)}{\sinh^2(a\pi \omega r_0)
+\cosh^2(\pi/2\sqrt{(4V_0a^2r_0^2-1)})}
\end{equation}
for $4V_0a^2r_0^2 > 1$. The P\"oschl-Teller potential
defined in equations (\ref{cosh1}) and (\ref{cosh2}) has been
used to extract quasi-normal modes of black holes, either
as an approximation \cite{Ferrari} or in obtaining exact results
in the Nariai spacetime \cite{Cardoso} for which the scalar field
equation reduces to the radial equation with the P\"oschl-Teller
potential.

\subsection{The rectangular barrier approximation}
In \cite{Matzners} Handler and Matzner used a rectangular
barrier as an approximate solution to obtain the transmission
coefficients corresponding to the black hole scattering 
problem of spin 1 particles. Their choice of the height of the barrier is $V_0$
with the same definition as explained above. The width $b$ 
is energy and $l$ dependent: $b=l/\omega$. The standard analytical results
for the rectangular barrier read for $(\omega r_0)^2 < r_0^2V_0$
\begin{equation} \label{cosh5}
|T_l|^2=\frac{1}{1+\frac{r_0^4V_0^2\sinh^2(y)}{4[r_0^4V_0\omega^2-\omega^4r_0^4]}}
\end{equation}
with $y\equiv l\sqrt{r_0^2V_0/\omega^2r_0^2 -1}$. 
For $r_0^2V_0 < (\omega r_0)^2$ one obtains
\begin{equation} \label{cosh6}
|T_l|^2=\frac{1}{1+\frac{r_0^4 V_0^2\sin^2(\tilde{y})}{4[\omega^4r_0^4-r_0^4V_0\omega^2]}}
\end{equation}
with $\tilde{y}\equiv l\sqrt{1-r_0^2V_0/\omega^2r_0^2}$.
Notice that with this prescription one cannot
calculate $T$ for $l=0$ which as far as the results of Handler and Matzner are concerned
is a valid assumption as they restrained themselves to $s=1$ and therefore via $l \ge s$ 
to $l > 0$.   
\subsection{The variable amplitude method}
In this section, 
$R_l(\omega)$ will be evaluated 
numerically via the variable amplitude method. 
The variable amplitude method was first introduced in \cite{tikochin} and
has been widely used to evaluate the reflection and transmission coefficients
for different potentials in literature \cite{rozkidunlee}.
This method 
involves writing the solution of the Schr\"odinger equation as
a superposition of the reflected and transmitted waves, namely,
$\psi_l(\omega,r_*) = T_l(\omega,r_*) [e^{i\omega r_*} + R_l(\omega,r_*) 
e^{-i\omega r_*}]$, which
leads to the following equation for $R_l$
\begin{equation}\label{Riccati1}
\frac{dR_l(\omega,r_*)}{dr_{*}}=-\frac{V(r_{*})}{2i\omega}\left[e^{i\omega
r_{*}}+R_l(\omega,r_{*})e^{-i\omega r_{*}}\right]^2 . 
\end{equation}
The absence of reflection behind the potential at $r_* \to - \infty$ imposes
the boundary condition $R_l(\omega, -\infty) = 0$ on the above equation.
The reflection
coefficient is given by $|R_l|^2 = |R_l(\omega,\infty)|^2$.
\subsection{Comparison of the methods and discussion of the results}
When we calculate the reflection coefficient, 
we expect it to be large and close to unity 
for energies much below the height of the barrier (where transmission is a
quantum mechanical possibility and hence very small).
Since transmission increases with energy, the reflection coefficient falls 
and at high energies (above the barrier) where transmission becomes 
the classical phenomenon and reflection a quantum mechanical effect, the 
reflection coefficient is negligibly small. 
In Fig. 3, we show the reflection coefficient as a function of energy for 
black hole scattering. On the left is shown the exact numerical result using the 
Regge-Wheeler potential and on the right the reflection coefficient evaluated 
from the expressions for $|R_l|^2 = 1- |T_l|^2$ discussed in the previous sections for 
a rectangular barrier and parametrized $\cosh^{-2}$ potential   
(which is similar in shape 
to a Gaussian barrier \cite{gausspot}).  
\begin{figure}[h]
\includegraphics[width=13cm,height=7cm]{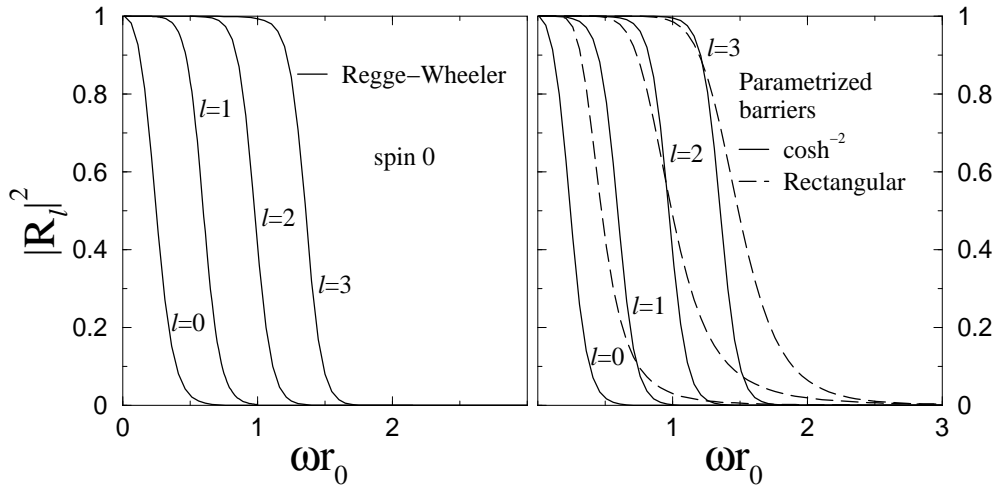}
\caption{\label{fig:eps3}
Reflection coefficient as a function of energy for $s=0$ and
for different values of 
$l$ in black hole scattering. On the left we plot
the numerical results and on the right the reflection coefficient for 
an adjusted rectangular barrier (see text) of height $V_0$ (dashed line) and for 
the parametrized $\cosh^{-2}$ potential (solid line). Since the results
for the latter almost coincide with the numerical ones we
plot them in two different boxes.}
\end{figure}
In Fig. 3 we have plotted the results in two separate boxes since the 
numerical results would almost
overlap with the results obtained from the $\cosh^{-2}$ potential.
This agreement is remarkable. In contrast to that, the results
obtained via the rectangular potential differ from the exact (numerical)
results.

\begin{figure}[h]
\includegraphics[width=13cm,height=11cm]{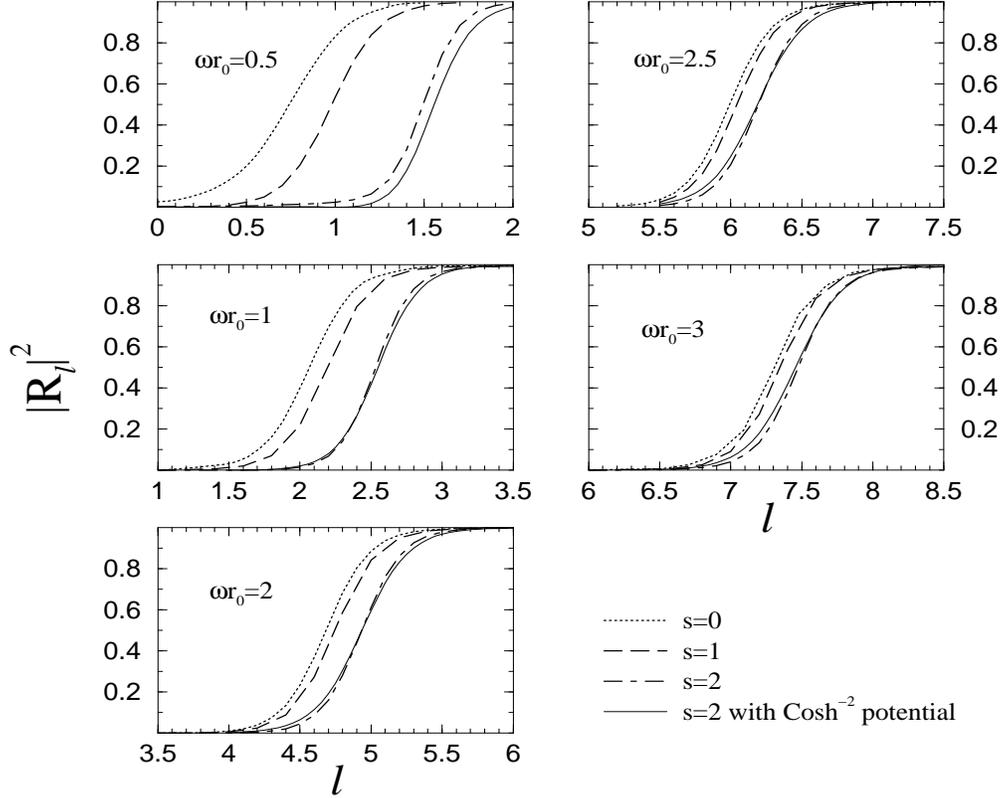}
\caption{\label{fig:eps4}
Reflection coefficient or the scattering amplitude squared 
as a function of $l$ for the
scattering of massless scalar ($s = 0$), electromagnetic ($s = 1$) and
gravitational ($s =2$) waves from a black hole. For comparison
we plot also for $s=2$ the results obtained from the parametrized 
$\cosh^{-2}$ potential. As expected the agreement with numerical results improves
with energy. For $s=1$ the same comparison is done in the next figure.
For $s=0$, we just mention that here the agreement between
the numerical and the semi-analytical results is the best among the three 
cases.} 
\end{figure}

In Fig. 4, we plot the numerically evaluated reflection coefficient as function of $l$ for 
different values of $\omega r_0$ and for different spins of the scattering 
particles. It is interesting to note that $|R_l|^2$ goes through a sudden rise 
at the critical values of $l$ listed in the tables and connected to
the orbiting phenomenon. 
The fact that the reflection coefficient for a given energy rises as a 
function of $l$ can be understood by examining the plot of the potential 
for different $l$ values at the same energy. In Fig. 2, we see that the effect 
of increasing $l$ is to increase the height of the barrier.
Hence, for example, an energy close to the top of the barrier for $l=3$ 
will lead to little reflection but at the same energy, the barrier for 
$l=7$ is much higher leading to larger reflection. 
For $s=2$ we made a comparison
with the corresponding $\cosh^{-2}$ potential. For small energies,
the agreement is still not perfect, but improves rapidly with
growing energies as is evident from the plots. This behaviour
is to be expected since the Regge-Wheeler
potential differs from the fitted $\cosh^{-2}$ case if the
energy of the particle is much below the height of the potential 
where the Regge-Wheeler displays an asymmetric tail. This small
mismatch between the the results obtained from
the two potentials should be also present for the cases $s=0, 1$, but 
will not be so prominent as for $s=2$. The reason is that 
$|R_l|^2$ for $s=0, 1$ saturates at smaller value of $l$. As a result
the difference between the small energy and the height is less
then in the case $s=2$ where the saturation sets in at higher $l$.
Indeed, as one can infer from Fig. 5 the agreement 
between the exact results for the Regge-Wheeler and the
$\cosh^{-2}$ potential is remarkably good for $s=1$ even for small energies.
\begin{figure}[h]
\includegraphics[width=13cm,height=11cm]{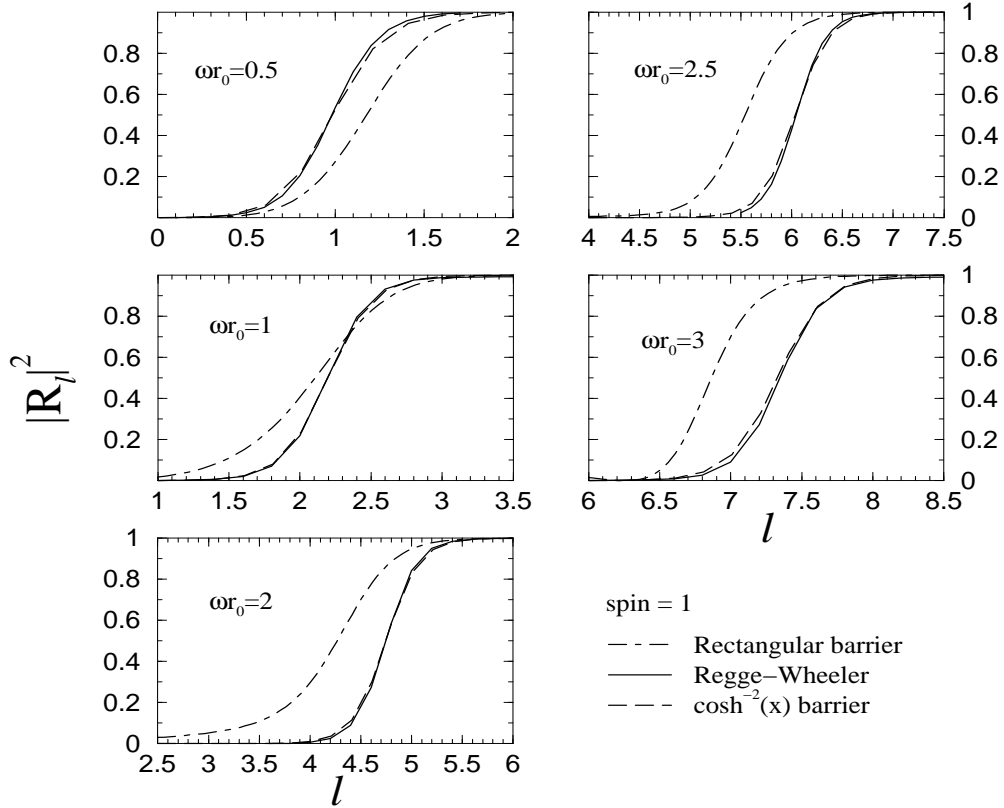}
\caption{\label{fig:eps5}
Comparison of the reflection coefficients calculated using numerical methods,
the adjusted rectangular barrier and the parametrized $\cosh^{-2}$ potential  
as a function of $l$ for the
case $s = 1$. Evidently the rectangular barrier (dash-dotted) is not a good approximation
whereas the $\cosh^{-2}$ (dashed lines) gives a very good agreement with exact numerical 
results (solid lines).}
\end{figure}
 
An approximate calculation of the magnitude of the reflection amplitude 
was done in the third reference in \cite{Matzners} (article 
by Handler and Matzner). For the case of $s = 1$, 
the authors approximated the potential by a rectangular 
barrier as explained before in the text 
and studied the features of the corresponding reflection amplitude 
as a function of the angular momentum $l$. The magnitude of the reflection 
amplitude for the various values of $\omega r_0$ studied here, 
started saturating to unity at a certain value of $l$ which the authors 
referred to as $l_g$, the critical value for the onset of glory scattering. 
It also showed a sudden rise through $1/2$ as a function of $l$ 
(this $l$ value however is different for the rectangular and realistic 
Regge-Wheeler case).
However, no interpretation was attempted to explain 
this fact. Indeed, here we have clearly connected it to the
orbiting effects.
The findings in \cite{Matzners} are not similar to those of the 
present work for the case $s =1$. A closer look at Fig. 5
reveals the differences between the exact results and the
results from a rectangular barrier. 
Not only is the shape of the reflection coefficients different,
but also the values at which 
the reflection coefficient makes a jump and at which it saturates to unity.

A possible explanation for $|R_{l_C}|^2 = 1/2$ as seen in Figs 4 and 5 for the 
numerical results 
can be found by examining the approximate expression 
of the reflection coefficient as obtained in the WKB approximation 
\cite{froeman}. In case of barrier penetration, when the energy $\omega^2$ 
of the incident particle lies below the top of the barrier, the 
semi-classical reflection coefficient is given as
\begin{equation}
|R_l|^2\, =\, {exp(2K_l) \over 1 + exp(2K_l)}
\end{equation}
with 
$$K_l = \int_{r_{1}}^{r_{2}} \, \sqrt{V(r_*) \, -\, \omega^2}\,dr_*\, , $$
where $r_{1}$ and $r_{2}$ are the classical turning points. For very small 
values of $\omega^2$, $|R_l|^2$ approaches unity. 
However, when $\omega^2$ equals the maximum height of the barrier, 
$r_1 \simeq r_2$, $K \to 0$ and $|R_l|^2 \to 1/2$. 
Thus one can relate the orbiting phenomenon with a critical $l$ value $l_C$ such 
that $|R_{l_C}|^2 = 1/2$. 
\subsection{Imaginary scattering phase shift}
Finally an interesting observation in connection with the orbiting is that the 
reflection coefficient which characterizes the critical value $l_C$ is 
related to the imaginary part of the scattering phase shift. If one relates 
the reflection amplitude to the $S$ matrix in 
three dimensional (3D) scattering, one can write 
$R_l(\omega) = exp(2 i \delta_l(\omega))$, where 
$\delta_l(\omega) = \delta_l^R(\omega)  + i \delta_l^I(\omega)$ 
in general, is the complex scattering phase shift. 
Thus, $R_l(\omega) = \eta_l(\omega)\, exp(2 i \delta_l^R(\omega)$, where, 
$\eta_l(\omega) = exp(-2 \delta_l^I(\omega))$ is known as 
the inelasticity parameter which can be less than or equal to 1. 
In the 1-dimensional case, 
the $S$ matrix is a $2 \times 2$ matrix with 
two channels, namely, transmission and reflection such that 
$|T|^2 + |R|^2 = 1$. 
If $\eta_l(\omega) =1$, it implies that $|R_l(\omega)|^2 = 1$ and there exists 
complete reflection. However, $\eta_l(\omega) = exp(-2 \delta_l^I(\omega)) < 1$ 
corresponds to the existence of the transmission channel. 
The reflection coefficient 
$|R_l|^2 = |\eta_l(\omega) \,exp(2 i \delta_l^R(\omega)|^2 
= \eta_l^2(\omega)$, i.e., $|R_l(\omega)|^2 = exp(-4\delta_l^I(\omega))$ 
and is related only to the imaginary 
part of the phase shift. The sudden rise in $|R_l(\omega)|^2$ 
at $l_C$ corresponds to a peak in $d|R_l|^2/dl$, where
\begin{equation}
{1 \over |R_l|^2} \, {d|R_l|^2\over dl}\, =\, - 4 \, {d\delta_l^I\over dl}\, .
\end{equation}
This should be contrasted with 
the semi-classical characterization in Eq. (\ref{deflphase}) where the 
orbiting and glory 
parameters are characterized using 
the real part of the scattering phase shifts. This is also due to the fact 
that 
the scattering phase shifts calculated within the semi-classical approaches 
such as the WKB are always real. Using an exact numerical evaluation of 
$|R_l|^2$ here, we find a connection of the orbiting parameters with the 
imaginary part of the phase shift.

\section{Re-interpretation of the Regge-Wheeler potential}
For $s = 1$ ($V_{eff} \propto V$), $r_C$ obtained 
from (\ref{veffgeo}) agrees with 
the quantum mechanical calculation via $(d^2|R_l|^2/dl^2)_{l=l_C} = 0$ 
(corresponding to the jump in $|R_l|^2$). 
The same would be true for $s = 0, 2$ if $r_C$ is computed 
through Eq.(\ref{veffgeo}), replacing 
therein $V_{eff}$ by $V$. The argument for $s = 0, 2$, namely, 
that $l_C$ is connected to the classical 
unstable circular orbit, can be now maintained if we attempt a 
reinterpretation of $V$. 
Restoring $\hbar$ in our expressions amounts to 
replacing $e^{i\omega t}$ by $e^{i(E/\hbar)t}$ or equivalently 
$\omega$ by $E/\hbar$.  
The Schr\"odinger-like equation then reads 
\begin{equation}\label{radialhbar}
\left[-\hbar^2 \frac{d^2}{dr_{*}^2} + \hbar^2 V(r)\right]\psi_{n \ell \omega
}=E^2\psi_{n \ell \omega } \, .
\end{equation}
Identifying $\hbar^2 l (l +1)$ with $L^2$, we can write 
\begin{eqnarray} 
\hbar^2 V &=& 2 \biggl [ \tilde{V}_{eff} + \hbar^2 {M (1 - s^2) \over r^3} 
\, \biggl ( 1 - {2M \over r} \biggr ) \, \biggr ]\, , \\ \nonumber
\tilde{V}_{eff} &=& \biggl ( 1 - {2M \over r} \biggr )\, {L^2 \over r^2} \, .
\end{eqnarray}
In identifying $\hbar^2 l (l+1)$ by $L^2$, we are going back from quantum mechanics to 
classical physics. 
It is reasonable to speculate that $\hbar^2 V$ represents 
the full effective potential, i.e., 
the classical part plus $\hbar^2$ quantum mechanical corrections. Indeed, 
$\hbar^2 V$ would then be the correct tool to 
calculate a classical unstable orbit.
Recently, $\hbar$ corrections to the Newtonian potential, 
have been discussed in \cite{newt}, where 
it was found that the additional terms are proportional to $\hbar M_1 M_2 G^2 / r^3$ ($G$ is 
the restored Newtonian constant). 
The procedure to arrive at 
such a result is to consider non-relativistic amplitudes with zero and 
higher order loop corrections. 
The difference from our case is that these corrections were calculated 
for massive particles where 
the Newtonian potential exists and the non-relativistic limit makes sense. 
In a massless case, such 
a procedure is not well-defined. 
Therefore our conjecture is well motivated but remains open. Since the dimensions 
of $V_{eff}$ from GR and $\tilde{V}_{eff}$ are different, let us be more specific. 
To make $\hbar^2 V$ dimensionless we divide it by the Planck mass squared, $E_{Pl}^2$ 
and identify $L^2/E_{Pl}^2 = \ell^2$. Thus, 
\begin{equation}
{\hbar^2 V \over E_{Pl}^2} \, =\, 2 \biggl [ \, V_{eff} \,+\, \hbar {2 G^2 M (1 -s^2) \over r^3} \, 
\biggl (\, 1 - {2GM \over r}\, \biggr) \, \biggr]\, .
\end{equation}  
Our speculation is simply to say that $V_{eff}$ 
receives a small quantum correction proportional 
to $\hbar$ in the above equation. This explains also the coincidences found in the 
previous section. Indeed, calculating $r_C$ from $V_{eff}$ or $V_{eff}$ plus quantum corrections
will give very similar results. Therefore it is not a surprise that 
$l_C^{QM}$ comes out quite 
close to the $l_C$ evaluated from $r_C$.

\section{Summary}
The behaviour of the reflection coefficient, $|R_l|^2$, 
which enters the scattering cross 
sections is investigated for the scattering of scalar, electromagnetic 
and gravitational waves from a Schwarzschild black hole. 
We paid special attention to the issue of orbiting effects
in a quantum mechanical scattering off black holes. 
Our investigation displays the following features: 
\begin{itemize}
\item[1.] For $s =0, 1,2$ we found that $|R_l|^2$ jumps at a certain critical 
value $l_C$, i.e. its second derivative with respect to $l$ is zero. 
The Regge-Wheeler potential is proportional to the 
classical effective potential for $s=1$ only. 
We find that $l_C$ is connected with the unstable 
circular orbit at $r_C = 3M$ for $s=1$. 
\item[2.] We find the $l_C$ values for $s = 0, 2$, too. 
Here also we would expect that the critical value $l_C$ is 
connected to an unstable circular orbit at a critical $r_C$. 
Interestingly, the $l_C$ values calculated using $V(r_C)$ lie
very close to those using $V(3M)$. 
This can be explained if we re-interpret the Regge Wheeler 
potential as the classical effective potential with $\hbar$ corrections. 
Notice that with the values of $l_C$ obtained via the rectangular 
potential, such a conclusion would be impossible as the jump occurs 
at a different $l_C$ and the connection to the unstable circular 
orbit is lost. 
\item[3.] For all values of $\omega r_0$ and spins $0, 1$ and $2$ we find 
that $|R_{l_C}|^2 = 1/2$. 
\item[4.] We have shown that the transmission and reflection 
coefficients of the potential proportional to 
$\cosh^{-2}(\alpha x -\beta)$  (P\"oschl-Teller) match
very well with the exact results. Since for this 
particular case the transmission coefficient
can be given analytically, this allows us to study the black hole scattering
in a semi-analytical way and supplements the conclusion
that such a potential is a good approximate tool in black hole physics
\cite{Ferrari, Cardoso}. 
Both the semi-analytical results and the 
numerical ones refine approximate results obtained elsewhere and reveal
some deficiencies of the approximation methods. 
\end{itemize}

\end{document}